\documentclass[preprint,aps,pra,preprintnumbers,superscriptaddress,floatfix]{revtex4}
\usepackage{graphicx,amsmath,amssymb,float,hyperref}

\begin{document}

\title{Ground State Quantum Coherences: from Quantum Beats to Strong Control.}

\author{Andres D. Cimmarusti}
\author{Burkley D. Patterson}
\author{Luis A. Orozco}
\affiliation{Joint Quantum Institute, Department of Physics and NIST, University of Maryland, College Park, MD 201742, USA.}
\author{Wanderson M. Pimenta$^1$}
\affiliation{Departamento de F{\'i}sica, Universidade Federal de Minas Gerais, Caixa Postal 702, Belo Horizonte, MG 30123-970, Brazil.}
\author{Pablo Barberis-Blostein}
\affiliation{Instituto de Investigaciones en Matem{\'a}ticas Aplicadas y en Sistemas, Universidad Nacional Aut{\'o}noma de M{\'e}xico, Ciudad Universitaria, 04510, M{\'e}xico, DF, Mexico.}
\author{Howard J. Carmichael}
\affiliation{Department of Physics, University of Auckland, Private Bag 92019, Auckland, New Zealand.}
\email{lorozco@umd.edu}

\begin{abstract}
Second order correlations reveal quantum beats from a coherent ground-state superposition on the undriven mode of a two-mode cavity QED system. Continuous drive induces decoherence due to Rayleigh scattering. We control this with feedback and explore postselection techniques to extract specific behavior.
\end{abstract}

\maketitle

\section{Introduction}

Our recent work in optical cavity QED has shown ground state quantum coherences generated by spontaneous emission \cite{norris10,norris12b}. These are a direct consequence of the internal structure of the rubidium atoms, and their response to small magnetic fields. Superpositions of ground state Zeeman sublevels undergo Larmor precession and manifest their frequency in quantum beats on the conditional intensity, the correlation function  $g^{(2)}(\tau)$. In contrast to recent experiments \cite{wilk07b,weber09}, which aim for deterministic quantum control, the ground-state coherence in our experiment is both prepared and read out by spontaneous emission. 

During these investigations \cite{norris12} we found that near resonant Rayleigh scattering is responsible both for an increase on the frequency of oscillation as well as a decrease in amplitude of the oscillations. This contribution presents our study of quantum feedback protocols to restore the coherence and control the oscillation frequency.  
The time scales of the quantum beats are such that current experimental and theoretical tools allow us to perform the feedback \cite{deutsch10,barberis10,norris11}.
  
Our previous experiments on quantum feedback in optical cavity QED \cite{smith02,reiner04a} also worked on conditional intensities with weak drives. We controlled the oscillatory exchange of excitation between the atomic polarization and the cavity mode. The frequencies were a factor of five higher than the decay rates of the system.  That protocol depended critically on the specific time of feedback application after a photon detection, as well as the electronic delay time. The protocols we are pursuing now rely on strong quantum feedback; we require a single photon detection to act on the system. This is in contrast with recent quantum control studies: Cavity QED experiments with Rydberg atoms in superconducting cavities \cite{sayrin11,zhou12} that used extensive calculations based on measurement outcomes to create and maintain a microwave Fock state; optical cavity QED experiments with two trapped Cs atoms to prepare a two-atom spin state using measurements of the cavity transmission with bayesian feedback control \cite{brakhane12}; and  quantum control of the full ground state manifold of Cs with off resonance optical polarimetry interrogation for measurement that point to future RF/microwave drives for more efficient actuators\cite{smith06,merkel08,mischuck12}.

\section{Our  cavity QED system}
We work with an optical cavity QED system in the intermediate coupling regime, where the single atom coupling to the excited state, $g/2\pi$=1.5 MHz, is comparable to the decay rates of the cavity, $\kappa =2\pi \times 3.0 \times 10^6$ s$^{-1}$, and the atomic excited state, $\gamma=2\pi \times6.07 \times 10^6$ s$^{-1}$. A detailed description of the apparatus is in Ref.~\cite{norris09a}. It consists of a 2 mm optical cavity, with 56 $\mu$m mode waist, whose two orthogonal polarization modes, resonant with the $D_2 ~~^{85}$Rb line of the cavity are degenerate to better than 0.2$\kappa$. 

We collect $^{85}$Rb atoms in a magneto optical trap (MOT) 7.5 cm above the cavity and direct them into the mode of the cavity by having an imbalance on the optical forces from a 1.5 mm hole on the vacuum-located retro-reflecting optics (quarter wave plate and mirror), an LVIS by the acronym given by its inventors in Ref.~ \cite{lu96}. The resulting continuous atomic beam has an average velocity around 15 m/s depending on the MOT parameters. The apparatus has the possibility to optically pump the atoms into the $F=3, m=0$ ground state before they enter the cavity. A small magnetic field (typically 5 Gauss) sets the quantization axis aligned with the polarization of the drive laser ($\pi$ polarization).

\label{sec:quant_beat}
\begin{figure}
\centering
  \includegraphics[width=0.8\linewidth]{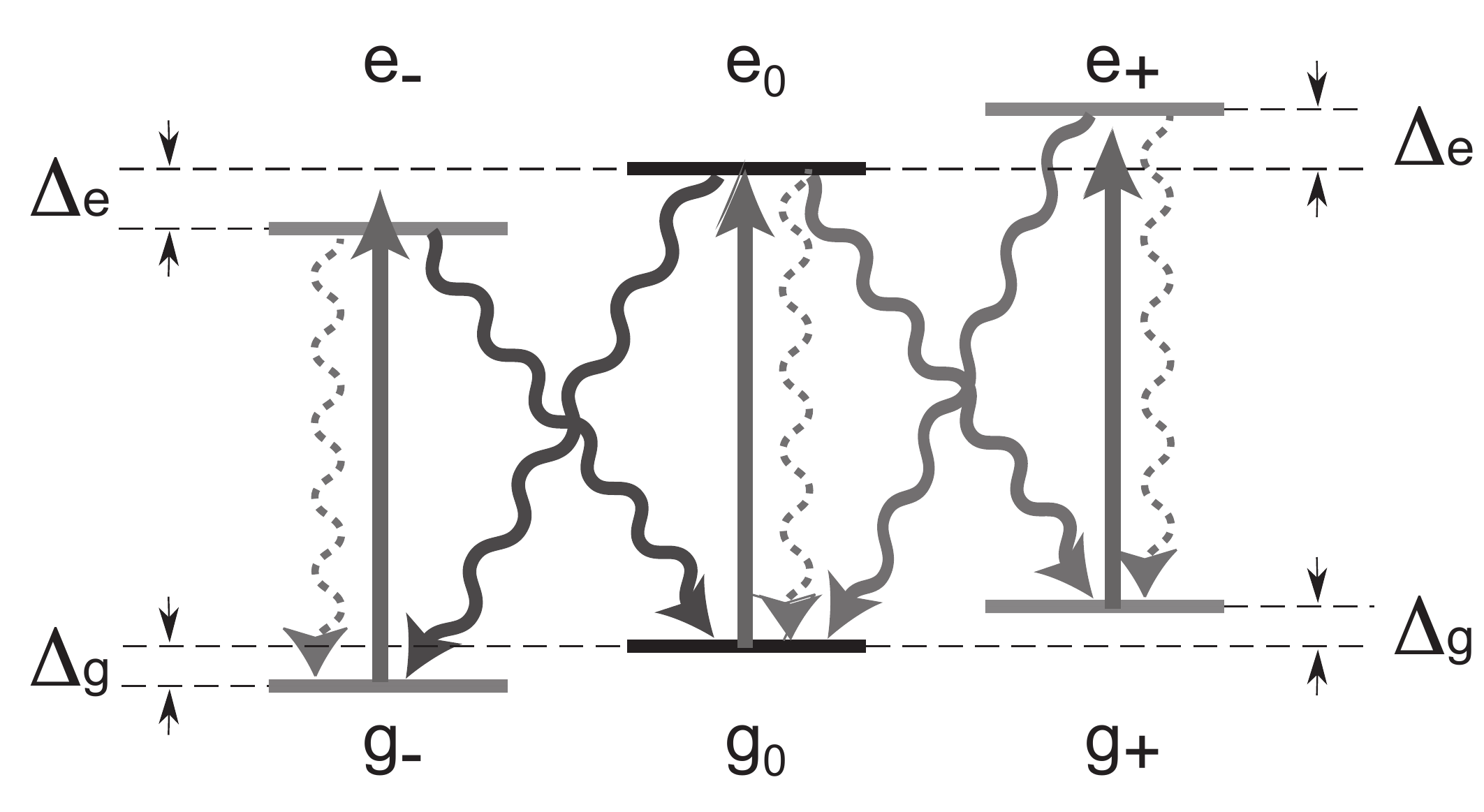}
\caption{Simplified atomic energy level structure of the $F=3 \rightarrow F=4$ $D_2$ line in $^{85}$Rb with Zeeman shifts.  Different magnetic $g$-factors yield $\Delta_{\rm e} > \Delta_{\rm g}$. Both the $\pi$ and $\sigma$ transitions are indicated. We consider only situations with a resonant $\pi$ drive. Adapted from \cite{cimmarusti13}.}
\label{fig:simplemodel}
\end{figure}

The atomic beam interacts with the two orthogonal modes of our high finesse optical cavity. The Zeeman structure of the $5S_{1/2}, F=3$ and the $5P_{3/2}, F=4$ states in $^{85}$Rb have different magnetic $g$-factors (see Fig.~\ref{fig:simplemodel}). The laser drives $\pi$ (V polarization) transitions, $F=3, m \rightarrow F=4, m$, (straight arrows in the figure). The spontaneous emission following atomic excitation can be through a circularly polarized photon (continuous undulating arrows) and they can transfer some of this energy to the orthogonal mode (H polarization). The result is that through the appropriate detection of an H 
photon, which cannot differentiate the origin of the spontaneous emission, we generate
a long-lived Zeeman superposition in the ground-state. Its signature is a quantum beat seen in the correlation function $g^{(2)}(\tau)$ of the mode with H polarization,  the undriven mode. If an atom enters the cavity in the $m=0$ ground state, the detection of a photon in the orthogonal mode sets the atom in a superposition of $m=\pm$ ground states. The prepared superposition then evolves in the magnetic field, acquiring a relative phase, until another $\pi$ excitation transfers the developed ground-state coherence to the excited state; subsequently, detecting a second (H polarized) photon projects the atom back into its starting state. The sequence overall realizes a quantum eraser \cite{scully82,zajonc83}, as the intermediate ground-state is not observed.

However, there can be several intervening $\pi$ spontaneous emissions (Rayleigh scattering) before the second H polarized photon---a $\sigma$ transition---can occur  \cite{uys10}. Each of these quantum jumps interrupts the atomic dipole and causes a small phase advance on the ground-state coherence.
The small phase shifts accumulate over time to become
a frequency shift~\cite{norris12}. The shift has decoherence from the randomness of the jumps that causes phase diffusion, which itself dephases the coherence. Fig. \ref{fig:simplemodel} does not include all the Zeeman sublevels of the ground state.

\section{Apparatus}
\label{sec:apparatus}
\begin{figure}
\centering
\includegraphics[width=0.8\linewidth]{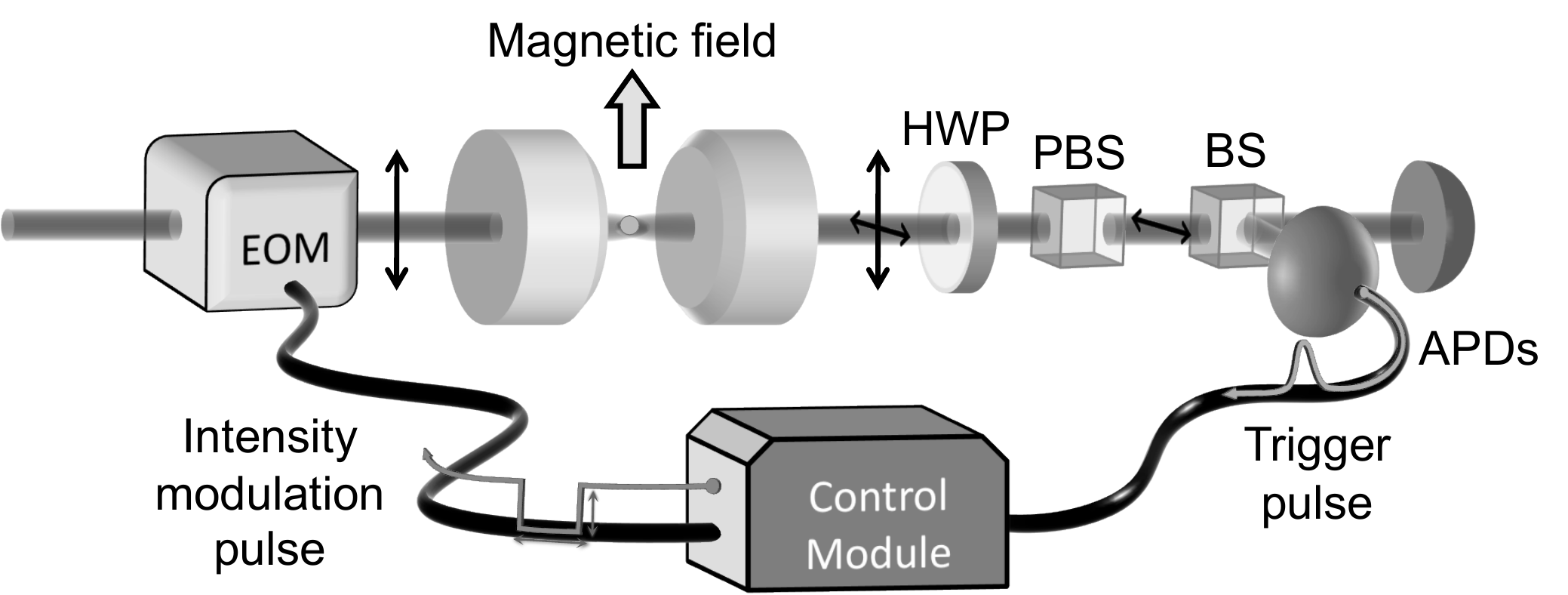}
\caption{Schematic of the experimental apparatus. The detection of a photon generates an electronic pulse that changes the amplitude of the laser drive for a pre-set amount of time. An electro optical modulator (EOM) sets the drive intensity. The light exits the cavity and passes through a half wave plate (HWP), a polarizing beam splitter (PBS), and a beam splitter (BS), which direct photons onto a pair of avalanche photodiodes (APDs). The photo-pulses from the APDs are correlated against the initial photon detection to obtain $g^{(2)}(\tau)$ (corrrelator not shown). Adapted from \cite{cimmarusti13}.}
 \label{fig:diagram}
\end{figure}

Figure \ref{fig:diagram} presents a schematic of the apparatus with the feedback loop included. The LVIS atomic source delivers, on average, a few maximally coupled atoms within the mode volume of the cavity at all times. The cavity supports two degenerate modes of orthogonal linear polarization (H and V). During their $5\mkern2mu\mu{\rm s}$ transit, the atoms interact with the orthogonally polarized modes and can spontaneously emit into the cavity. We drive the cavity QED system resonantly or a few linewidths away from the $5S_{1/2}, F=3, m=0 \rightarrow 5P_{3/2}, F=4, m=0$ transition of the $D_2$ line of $^{85}$Rb.

We use the following procedure with the apparatus on Fig. \ref{fig:diagram} for the results that we present here in Sec.~\ref{preliminary}.
The light at $780\mkern2mu{\rm nm}$ passes through an EOSpace fiber electro-optic modulator before reaching the cavity. This device generates amplitude modulation sidebands  on the light. The upper sideband drives the system. The setup allows us to rapidly manipulate the amplitude of the drive. We use a digital delay generator  (DG) to generate a fast electronic pulse, variable in time, that opens an RF attenuator, to feed the EOM. The TTL pulse out of the APDs travels to a correlator card, where each pulse arrival is time-stamped and stored. We split one of the APD TTL pulses before reaching the correlator card and use it to trigger the DG, which then drives the attenuator. 

We detect the coherence, oscillating directly at the Larmor frequency, using an interference process between the signal generated by the spontaneous emission and  a small part of the light exiting the drive mode using a half wave plate (HWP in Fig.~\ref{fig:diagram}).

The decoherence of the superposition is  intimately related the drive laser \cite{norris12}. The protocol to correct for this consists of reducing the amplitude of the $\pi$ drive to the cavity after a pre-set time from the detection of the first photon, and then returning the amplitude to the previous level after a fixed time to look at the oscillations. Since the frequency of oscillation is, to first order, set by the Larmor precession frequency; the atoms preserve the phase without interrogation, continuing their oscillation in the ground state. For more details see \cite{cimmarusti13}

\section{Theory}
\label{sec:theory}
A $\pi$ polarized coherent field with amplitude $\alpha$, resonant with the transition $|g_0\rangle\rightarrow|e_0\rangle$, drives $\pi$ transitions on the vertical mode of a cavity QED system in the presence of a magnetic field (see Fig.~\ref{fig:simplemodel}). The drive of the cavity is on one of the two orthogonal polarization modes ($\pi$ in Fig.~\ref{fig:simplemodel}, also identified as $V$) and the detection is in the other. Two shifts compete in this process, one is the AC Stark shift present because there is a field driving the system $\alpha$, but a second one is a shift that comes from the phase advance that appears every time there a quantum jump, Rayleigh scattering.The spontaneous emission falls into the cavity mode and can have an $H$ polarization component. The detection at the output of the cavity (see Fig. \ref{fig:diagram}) is a correlator that looks on the $H$ mode. We can have spontaneous emission mixed with a local oscillator of strength $\epsilon$. The correlator performs a conditional intensity measurement so that the detection of a first $H$ photon prepares a ground-state superposition \cite{norris10} that evolves in time as
\begin{equation}
  |\psi_{\rm g}(t)\rangle = C_0\frac{1}{\sqrt{2}}(e^{i (\Delta_{\rm g}+\Delta_{\rm AC})t}|g_-\rangle + e^{-i (\Delta_{\rm g}+\Delta_{\rm AC})t}|g_+\rangle )
  +C_1 |g_0\rangle
\label{eq:groundstate}
\end{equation}
where the amplitudes $C_0$ and $C_1$ depend on the strength and phase of the local oscillator. 
To lowest order in $g^2|\alpha|^2$ (drive intensity), with $\Delta=\Delta_{\rm e}-\Delta_{\rm g}$ the difference of the excited and ground state Zeeman shifts (see Fig.~\ref{fig:simplemodel}), the ground-state AC Stark shifts are
\begin{equation}
  \Delta_{\rm AC}= - \frac{g^2|\alpha|^2\Delta}{(\gamma/2)^2+\Delta^2}
\label{eq:ACStarkshift}
\end{equation}
for state $|g_+ \rangle$ and $-\Delta_{\rm AC}$ for $|g_- \rangle$.

The amplitudes in Eq.~(\ref{eq:groundstate}) couple to the corresponding excited-state amplitudes, driving a steady-state superposition:
\begin{equation}
  \label{eq:excited-jump}
  |\psi_{\rm e}(t)\rangle = C_0\frac{g \alpha}{\sqrt{2}}\left(\frac{e^{i(\Delta_{\rm g}+\Delta_{\rm AC})t}}{\gamma/2 - i \Delta} |e_-\rangle
    + \frac{e^{-i (\Delta_{\rm g}+\Delta_{\rm AC})t}}{\gamma/2 + i \Delta} |e_+\rangle\right)+C_1\frac{g\alpha}{\gamma/2}|e_0\rangle.
\end{equation}
The excited-state amplitudes follow the ground-state oscillation; the excited-state splitting enters through the factors $\gamma/2\pm i \Delta$ only, which carry a phase shift. Crucial for this effect is the difference in the Larmor precession frequency (g factors) of the ground and excited states of the atom.

Rayleigh scattering, that goes undetected with the correlator, with
jump rate $\Gamma=2g^2|\alpha|^2/(\gamma/2)$, turns off  the driven dipole between ground and excited states and the amplitudes of Eq.~(\ref{eq:excited-jump}) are transferred to the ground state. It follows that each time a quantum jump occurs there is a phase advance; if $n$ quantum jumps occur, Eq.~(\ref{eq:groundstate}) is replaced by:

\begin{eqnarray}
\mkern-60mu{\cal{N}}_n|\psi_g(t)\rangle&=& \frac{C_0(\gamma/2)^n}{\sqrt2}\left\{\frac{(\gamma/2+i\Delta)^n}{[(\gamma/2)^2 + \Delta^2]^{n/2}} e^{i(\Delta_g+\Delta_{AC})t}|g_-\rangle\right.\nonumber\\
&&\left.+\frac{(\gamma/2-i\Delta)^n}{[(\gamma/2)^2+\Delta^2]^{n/2}}e^{-i(\Delta_g+\Delta_{AC})t}
|g_+\rangle\right\}+C_1[(\gamma/2)^2 + \Delta^2]^{n/2}|g_0\rangle\, ,
\label{ground-jump}
\end{eqnarray}
with normalization factor:
\begin{equation}
{\cal{N}}_n=\sqrt{|C_0|^2(\gamma/2)^{2n}+|C_1|^2[(\gamma/2)^2+\Delta^2]^n}.
\end{equation}
The ground-state superposition acquires a phase advance. If the number of quantum jumps increases with time, there can be a phase shift but also there is phase diffusion that causes decoherence of the ground state superposition. 

Averaging against a Poisson distribution with mean $\Gamma t$ the frequency shift to first order in $2\Delta/\gamma$ gives:

\begin{equation}
\Delta_{\rm jump} = \Gamma \frac{2\Delta}{\gamma}=\frac{8 g^2|\alpha|^2\Delta}{\gamma^2}=-2\Delta_{\rm AC}.
\label{eq:jumpshift}
\end{equation}
These terms represent an additional frequency shift arising from the mean rate of phase accumulation from quantum jumps due to Rayleigh scattering. For the $(g_\pm,g_0)$-coherence, the net differential ground-state shift, in the low drive limit, becomes:
\begin{equation}
\Delta_{\rm light}=(\Delta_{\rm AC}+\Delta_{\rm jump})\approx-\Delta_{\rm AC},
\label{eq:deltalight}
\end{equation}
and $2\Delta_{\rm light}=-2\Delta_{\rm AC}$ for the $(g_+,g_-)$-coherence.

Also the related damping term, which decoheres the quantum beats at a rate (to lowest order in $2\Delta/\gamma$) is:
\begin{equation}
\Gamma_{\rm decoh} = \Gamma \frac{\Delta^2}{(\gamma/2)^2}=2g^2|\alpha|^2\frac{\Delta^2}{(\gamma/2)^3}.
  \label{eq:jumpwidth}
\end{equation}
The decoherence arises from the phase diffusion which accompanies the average phase drift responsible for the frequency shift. The two aspects, drift and diffusion, come together as a package from the stochastic nature of the jump process.

Both the frequency shift (Eq.~\ref{eq:jumpshift}) and the damping rate (Eq.!\ref{eq:jumpwidth}) depend on the intensity of the drive $|\alpha|^2$ and on the detuning (magnetic field through the different Zeeman shifts in the ground and excited states). The effect is present in a multilevel atom and persists for small detunings of the order of the excited state linewidth $\gamma$. Recent work in ion traps \cite{uys10}, has observed the enhanced decoherence from Rayleigh scattering, but their excitation scheme has a large detuning making each quantum jump contribute a large phase shift.

\section{Results}
\label{preliminary}
\begin{figure}[h]
\centering
\includegraphics[width=0.8\linewidth]{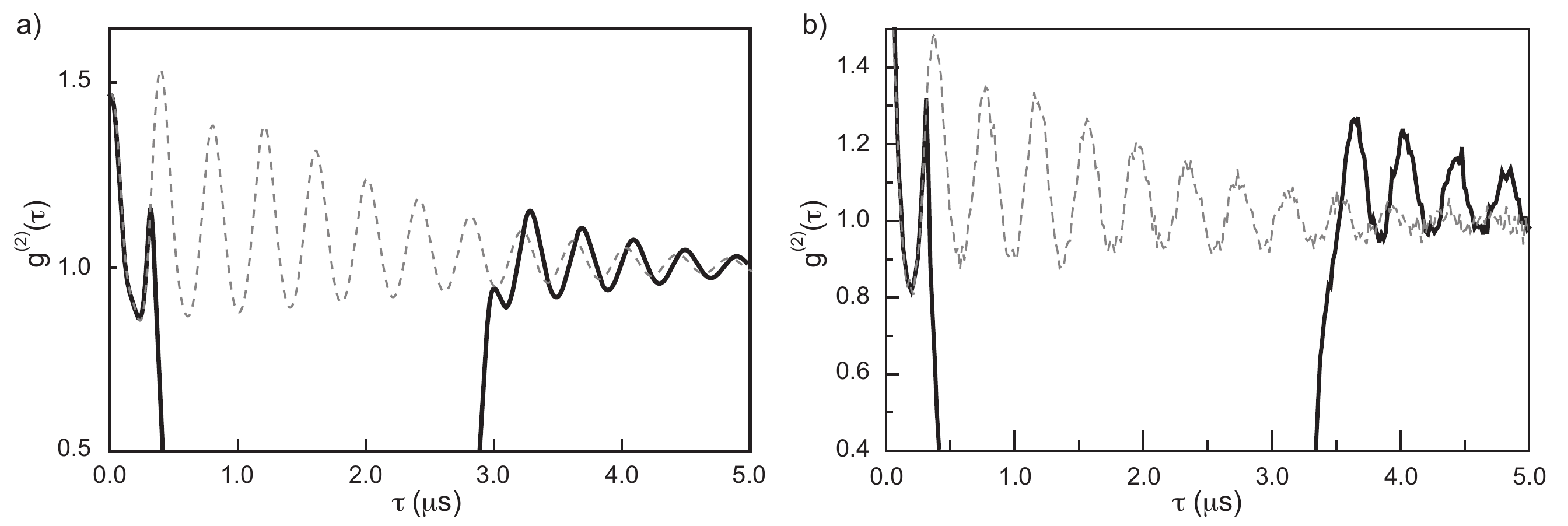}
\caption{a) Theoretical simulation for the conditional intensity, $g^{(2)}(\tau)$, without feedback (dashed) and with feedback (continuous). b) Measured conditional intensity, $g^{(2)}(\tau)$, with fixed feedback pulse length ($3\mkern2mu\mu{\rm s}$) and two amplitudes: no feedback (dashed) and with the intensity reduced to 5\% of the total (continuous).}
 \label{fig:data1}
\end{figure}
 Figure~\ref{fig:data1}a shows the results of a simulation of the experiment with a quantum Montecarlo method  \cite{carmichael93book}. To reduce the decoherence process due to Rayleigh scattering we create and capture the coherence, preserving it in the dark, where it evolves at the ground state Larmor frequency without interrogation, and then we drive it again to measure it. The phase difference visible after the oscillations return is a measure of
the average number of intervening quantum jumps. 
The value of the driving field $\alpha$ determines the frequency shift (Eq.~\ref{eq:deltalight}) and decoherence rate (Eq.\ref{eq:jumpwidth} induced by Rayleigh scattering. The dashed gray line shows the correlation function with no control. The decay is clearly observable. The continuous black trace shows the results when the drive is fully turned off for 2.5 $\mu$s. When the oscillation returns there is both an amplitude change and a phase shift. All the parameters in the simulation are within the allowed ranges of the experiment.

Figure~\ref{fig:data1}b shows also two traces. The dashed line is the measured result for the case with no feedback. The oscillations decay in about 3 $\mu$s, in contrast the continuous black line  shows results with a fixed feedback pulse -- reduce the $V$ drive on the cavity to 5\% -- for 3 $\mu$s. This allows us to observe the large change in the size of the oscillation and the phase shift after light is back on, as measured by the conditional intensity $g^{(2)}(\tau)$. 

The subtle effect of Rayleigh scattering on the conditional ground state quantum beats that increases their frequency and decreases their amplitude is possible to correct. We have performed simple quantum feedback following the detection of a first photon, that heralds the creation of the superposition. We turn off the drive and let the coherence evolve in the dark. We avoid the quantum jumps that although small enough individually accumulate to produce measurable frequency shifts accompanied by faster decoherence in the amplitude
though the induced cumulative phase diffusion.

\section{Post Selection}

\begin{figure}[t]
\centering
\includegraphics[width=0.8\linewidth]{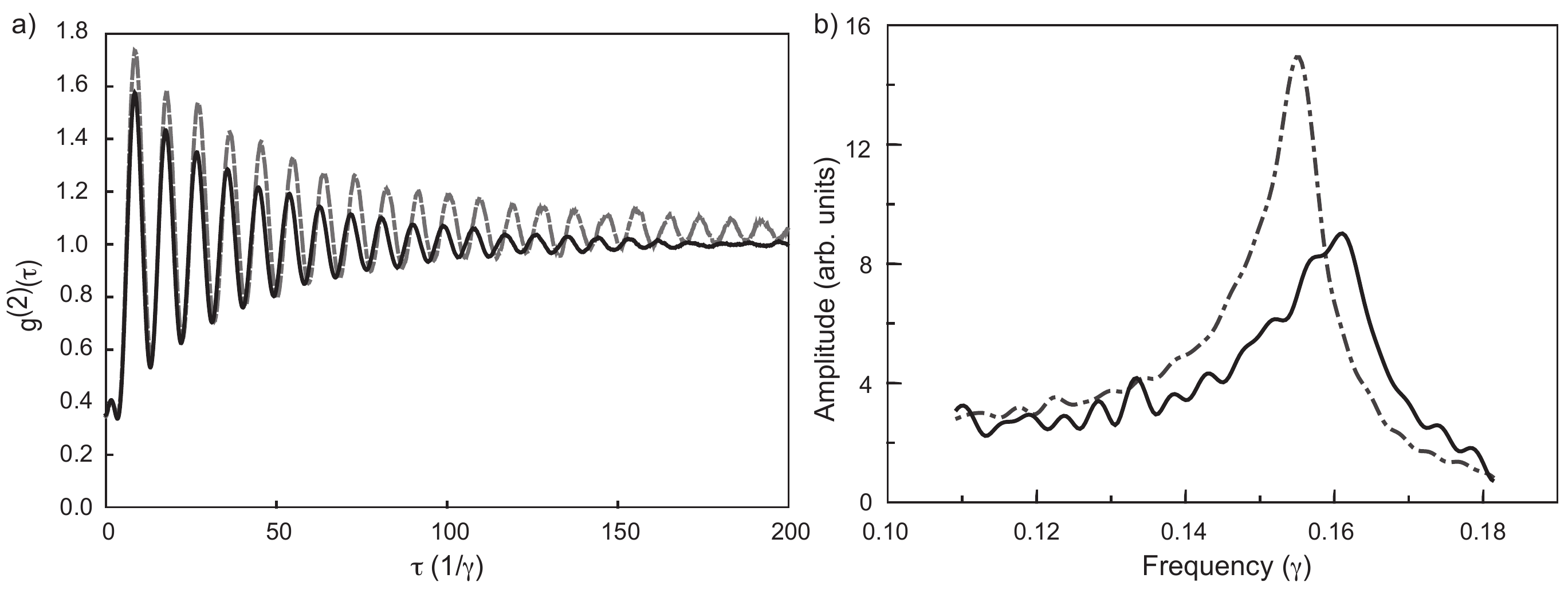}
\caption{a) Simulations and b) FFT of  simulations of the  correlation functions of the undriven mode with 0.55 photons in  the driven mode. The black continuous traces are obtained without any  post selection on the number of spontaneous emissions; the gray dashed traces have post selection of less than 14 spontaneous emission events in 300 atomic lifetimes ($1/\gamma$).}
 \label{fig:post1}
\end{figure}
Post selection is a powerful approach in quantum optics and quantum information. It forms now part of many protocols and we  present here an example that addresses decoherence and frequency shifts of the conditional quantum beats. We use quantum Montecarlo techniques, basically the same program that produces results as those in the experimental simulation of Fig.~\ref{fig:data1}(a).  

Figure \ref{fig:post1} shows the $g^{(2)}\tau$ of the undriven mode of our system. The small antibunching observed at $\tau=0$is the result of quantum simulations for a fixed maximally coupled atom. 
It is possible to track in the simulation the events of spontaneous emission to the side of the mode. Part a) of the figure shows three correlation functions created based on the long time record of H emissions out of the cavity, but also tracking the spontaneous emission events. The continuous black trace is without any post selection, it shows the average of all possible events. If we set a maximum of 14 spontaneous emission events during the 300 atomic lifetimes, after the initial detection on the undriven mode, we get the dashed grey curve. This clearly shows less decoherence, and the frequency of oscillation is smaller, which indicates a lower number of phase shifts.  Part b) of the figure shows the FFT using the same color code as in part a, they show a clear frequency change and their widths do change. The implementation of these protocols with data from an experiment will be complicated, but we are looking into a few possibilities.

These are preliminary calculations, but they are encouraging enough that we are setting the extra detection port for the driven mode on the upgraded apparatus. This will allow us to time stamp all the photons escaping the driven mode and then perform the post selection on the full records.

\begin{figure}
\centering
\includegraphics[width=0.8\linewidth]{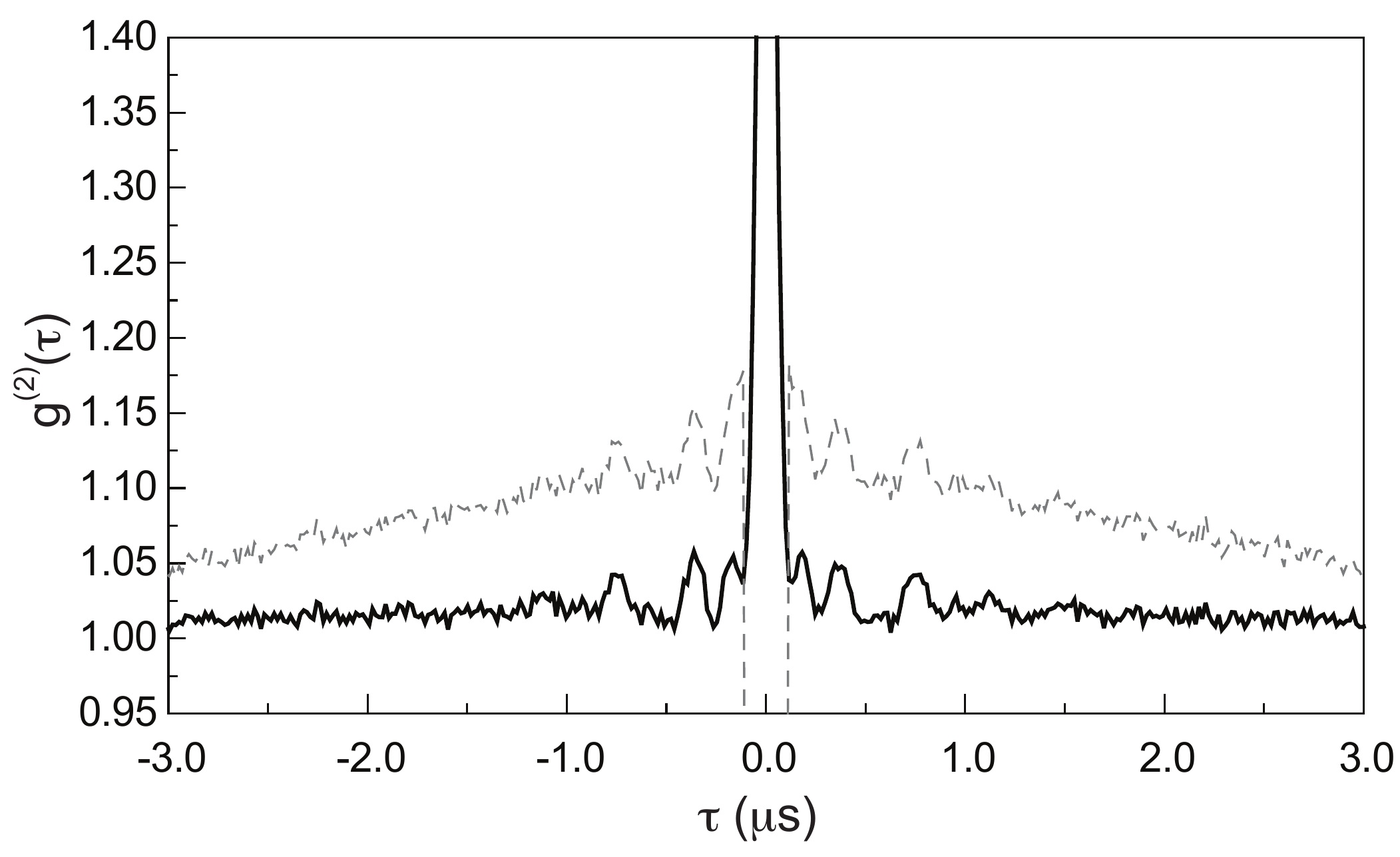}[h]
\caption{Conditional intensity with $N_{eff}=1.5$, with no post selection (black line) with 100 ns conditioning window (gray dashed line).}
 \label{fig:post3}
\end{figure}

We recently modified our $g^{2}$ analysis program to include a
time-filtering routine. This new process checks the time of
arrival of the next photon. If this time is shorter than a specified
window, then this photon as well as the previous one are not
considered in the construction of the correlation function. When such an
event is detected, the program skips a set time worth of photon
clicks.

The rationale behind this kind of ``high-pass time'' filter is that if
we are interested only in extracting from the data the photon arrivals
when only one atom is in the cavity, then we must choose a window of
the order of the excited state lifetime (~26 ns). If there is only one atom, there should be antibunching in the cavity at $\tau=0$ \cite{mielke98,foster00pra, terraciano09}.

We have realized the data from Ref.~\cite{norris12b} when we were sure we operated with a large number of atoms. Characterize this by an effective number of atoms ($N_{eff}$) \cite{carmichael99} that given our parameters means that on average there are seven randomly distributed atoms on the cavity. The signature of having only one atom in the cavity is a modification of the correlation function that shows the oscillations on top of a Gaussian looking background that comes from the transit time of the atom across the cavity mode \cite{norris10, norris12b}.

Figure \ref{fig:post3} shows preliminary results of the correlation function showing quantum beats. The black line has no post selection while the gray dashed line explicitly deletes a region of 5 $\mu$s after two events happen on a window of 100 ns. This is the antibunching enforcement. It is clear that the single atom signature is visible.  

We continue to improve the filtering algorithm and expect to be able to look at what remains in the correlation function around $\tau=0$. The results are encouraging.

\section{Conclusions}
Spontaneous emission prepares quantum beats
from the ground state. Quantum beats have sufficiently long coherence to interrogate and study.
However, too much spontaneous emission can
destroy the quantum beats creating frequency shifts and
decoherence. The origin of this shifts and broadening has to do with the  phase jumps associated with the unobserved Rayleigh scattering in our system.

The implementation of strong feedback control significantly decreases the decoherence and frequency shifts. The protocol consists of turning off the drive to the cavity QED system after the first photon detection let the superposition evolve in
the dark. This enhances the lifetime of the coherence and also improves the visibility of the conditional quantum beats.

Post selection of data shows changes in the
correlation function consistent with one photon.
Simulations show other avenues for post selection protocols.

\section{Acknowledgments}
We would like to thank the help over the years that D. G. Norris, J. A. Crawford, and C. D. Schroeder have given us in the experiment. Funding for this has come from the National Science Foundation of the United States, CONACYT from Mexico, and the Mardsen Fund of the Royal Society of New Zealand.

%

\end{document}